# Support-Area Dependence of Vibration-Insensitive Optical Cavities

Won-Kyu Lee[1,2,*], Sang Eon Park[1,2], Chang Yong Park[1,2],

Dai-Hyuk Yu[1], Myoung-Sun Heo[1], and Huidong Kim[1]

[1]*Korea Research Institute of Standards and Science, Daejeon 34113, Korea*
[2]*University of Science and Technology, Yuseong, Daejeon 34113, Korea*
*Corresponding author: <u>wonqlee@kriss.re.kr</u>

The vibration sensitivities of optical cavities depending on the support area were investigated, both numerically and experimentally. We performed numerical simulations with two models: one with total constraint of the support area, and the other with only vertical constraint. An optimal support condition insensitive to the support's area could be found by numerical simulation. The support area was determined in the experiment by a Viton rubber pad. The vertical, transverse, and longitudinal vibration sensitivities were measured experimentally. The experimental result agreed with the numerical simulation of a sliding model (only vertical constraint).

## I. INTRODUCTION

Narrow-linewidth lasers are essential tools for various applications, including optical clocks [1-4], low-phase-noise microwave generation [5], fundamental tests of relativity [6], gravitational-wave detection [7], and dark matter searching [8, 9]. The laser linewidth is narrowed usually by stabilizing its frequency on an ultrastable optical cavity resonance. Because such an optical cavity is sensitive to external vibrational noise, there have been a number of studies during the last decade on support methods to minimize the sensitivity of both horizontal cavities [10-14] and vertical ones [15, 16], and the frequency stability of lasers reached the thermal noise limit, which is given by the Brownian noise from mirror coatings, mirror substrates, and cavity spacers [12, 16]. To decrease the thermal-noise limit further [17], researches are being actively undertaken with longer cavities [17-21], with cavities at cryogenic temperatures [22, 23], or using new materials with lower mechanical loss [24]. However, longer cavities are generally more sensitive to vibrational noise, and their frequency stabilities are still limited by seismic noise [21]. Thus, more works are required in reducing the vibration sensitivity of optical cavities.

For the clock lasers of the Yb optical lattice clock at KRISS (Korea Research Institute of Standards and Science) [25-27], cutout-type vibration-insensitive horizontal optical cavities were adopted [11, 12]. Optimal cavity support positions with the smallest vibration sensitivity can be determined by numerical simulations before the experiments [10-24]. Optical cavities are usually supported by small Viton rubber balls [11, 12, 18, 19, 21]; however, considering that rubber balls can be squeezed, the numerical simulations have limitations in this case, because the support-area sizes are ambiguous and the pressure distribution over the support area is not uniform. With this concern, we used Viton pads instead of balls to support the optical cavity. We performed finite element analysis (FEA) for the vibration sensitivity of our cavities, varying the support's position and its diameter with a fixed support point (fully constrained) and a sliding support (only vertically constrained). Also, an optimal support condition, insensitive to support area, could be found. We compared these numerical simulations with experimental results.

## II. CAVITY DESIGN AND FINITE ELEMENT ANALYSIS

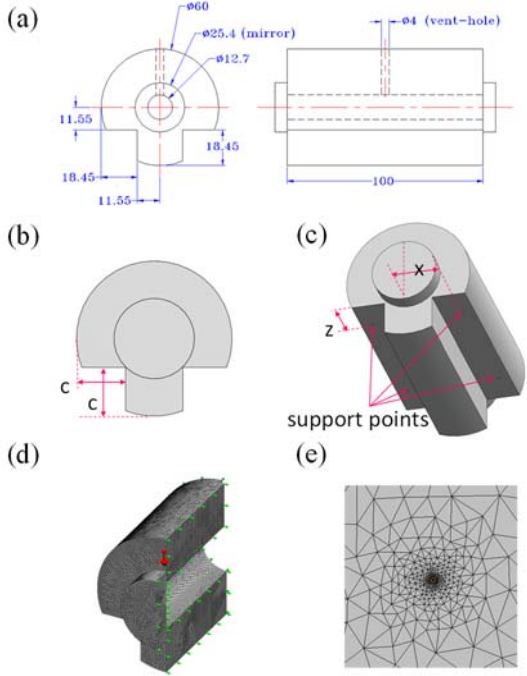
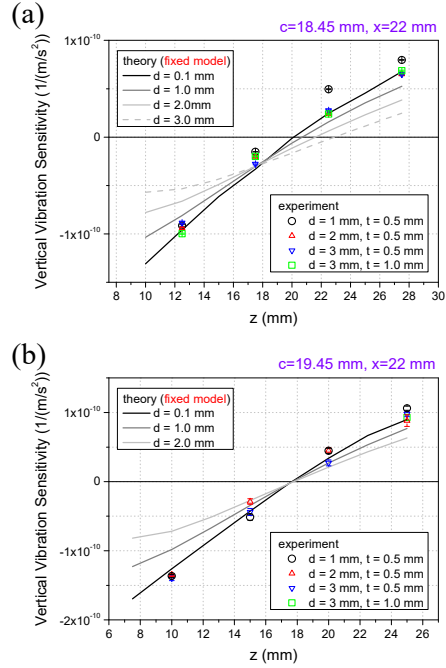

Figure 1. (a) The schematics of the cavity structure (left, front view; right, side view). Dimensions are given in mm. (b) Front view of the cavity with a cutout depth of $c$. (c) Positions of the four support points, with a transverse displacement $x$ and a longitudinal displacement $z$. (d) Numerical simulation of cavity deformation with a quarter section of the cavity, utilizing symmetry (e) Meshes around a support point in the finite element analysis.

Figure 2. (a) Dependence of the vertical vibration sensitivity of Cavity1 on $z$ and on the support point's diameter $d$, which was calculated by FEA with fixed supports. The values of $c$ and $x$ were fixed at 18.45 and 22 mm respectively. The experimentally measured values are also plotted for various $d$ (support diameter) and $t$ (Viton pad thickness). (b) The dependence of Cavity2 under the same conditions as in (a), except for the change of $c$ to 19.45 mm.

The shape of the cavity (front and side views, with dimensions in mm) in this research is shown in Figs. 1(a) and 1(b), with the definition of the cutout depth $c$. We used two cavities, differing only in cutout depth: 18.45 mm for Cavity1 and 19.45 mm for Cavity2. The positions of the support points are shown in Fig. 1(c), with transverse displacement $x$ and longitudinal displacement $z$. The length and outer diameter of the cavity spacer are 100 mm and 60 mm, respectively. The spacer has a bore with a diameter of 12.5 mm, and a vertical vent hole in the middle with a diameter of 4 mm. The cavity is formed by a plane mirror and a concave mirror with a radius of curvature of 500 mm. The diameter and thickness of each mirror are 25.4 mm and 6.2 mm respectively. The cavity spacer and mirror substrates are made of ultralow-expansion glass (ULE® [28]). With multilayer dielectric coating, the finesse values were 160,000 and 300,000 for Cavity1 and Cavity2 respectively, at 578 nm. The mirror substrates have antireflective coatings on their outer surfaces.

The locations of the support points have to be optimized to minimize the vibration sensitivities [10-24]. Thus, we performed finite element analysis (FEA) to estimate the optimal positions of support points by varying $c$, $x$, and $z$ [29]. Considering the symmetry, we simulated only a quarter section of the cavity (Fig. 1(d)), to reduce computation time. We also performed FEA for various values of the diameter $d$ of the support point. The areas of the support points were defined by those of the Viton pads in our experiment. Two kinds of simulation model were adopted: one with a fixed model (the support areas were totally fixed), and the other with a sliding model (the support areas were only vertically fixed). The material parameters of ULE used in the FEA were a Poisson's ratio of 0.17, the density of 2.21 g/cm$^3$, and an elastic modulus of 67.6 GPa [28]. Typical meshes in the FEA are shown in Fig. 1(e), the maximum sizes of which were 0.1 mm for the support areas and highly-reflective-coated mirror surfaces, where high spatial resolution is required, and 1 mm for other parts of the cavity. The FEA was performed statically, because seismic noise of higher frequency would be attenuated by vibration isolators in the experiment. The change of cavity length was obtained by calculating the distance between the center points of the two mirror surfaces when static vertical acceleration of 1 $g$ (9.8 m/s$^2$) was applied.

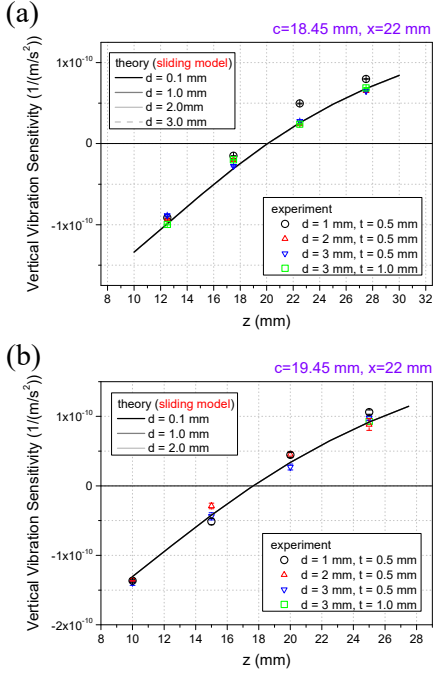

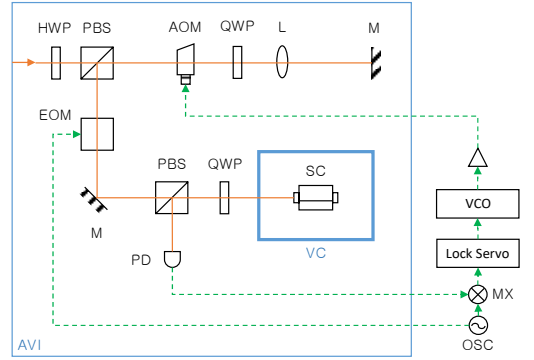

Figure 4. Experimental setup for measuring vibration sensitivity: AVI, active vibration isolator; HWP, half-wave plate; PBS, polarizing beam splitter; AOM, acousto-optic modulator; L, lens; M, mirror; EOM, electro-optic modulator; QWP, quarter-wave plate; PD, photodiode; SC, supercavity; VC, vacuum chamber; VCO, voltage-controlled oscillator; MX, mixer; OSC, two-channel function generator.

Figure 3. (a) Dependence of the vertical vibration sensitivity of Cavity1, calculated by FEA with sliding supports. The values of $c$ and $x$ were fixed at 18.45 and 22 mm respectively. The experimentally measured values are also plotted, for various $d$ (support diameter) and $t$ (Viton pad thickness). (b) The dependence of Cavity2 under the same conditions as in (a), except for the change of $c$ to 19.45 mm.

In Fig. 2(a), the FEA result for Cavity1 with the fixed model is shown, for varying $d$ and $z$. The vibration sensitivity (the normalized change in cavity length divided by the acceleration) depending on $z$ is shown for $d$ values of 0.1, 1.0, 2.0, and 3.0 mm. The values of $c$ and $x$ were fixed at 18.45 mm and 22 mm respectively. We found that the $z$ positions for zero longitudinal displacement (*i.e.* minimum vibration sensitivity) are different for various support areas. There exists only one set of parameters giving minimum vibration sensitivity for various diameters of the support area. We also found a value of $c$ for which the $z$ values for minimum vibration sensitivity are the same, regardless of the support area. When the value of $c$ was changed to 19.45 mm with the same values of the other parameters, a support-area-independent support condition was found at $z = 17.5$ mm, as in Fig. 2(b), on the basis of which Cavity2 was designed. As stated in the introduction, when rubber balls are used for the cavity support, there is ambiguity regarding the support area. This support-area-independent condition is expected to be useful in that situation. It is notable that the smaller the support area, the larger the slope (the vibration sensitivity per longitudinal displacement), for both Cavity1 and Cavity2.

The dependence of the vertical vibration sensitivity on $z$ for Cavity1 and Cavity2, calculated by the FEA with the sliding model, is shown in Figs. 3(a) and 3(b) respectively. In this case there was no dependence of the vibration sensitivity on support area, and the $z$ dependence was approximately the same as that for a very small support area with the fixed model ($d = 0.1$ mm, in Fig. 2). This result seems reasonable, because the sliding model can be considered to be the small-area-limit of the fixed model. The $z$ values for zero longitudinal displacement were 20.0 and 17.5 mm for Cavity1 and Cavity2 respectively.

## III. EXPERIMENTAL SETUP

Figure 4 shows the experimental setup for measuring vibration sensitivity. The laser has a wavelength of 578 nm [26, 30, 31], which is a probe laser for the Yb lattice clock transition. The frequency of the input laser in Fig. 4 was stabilized in advance to a resonance of another cavity, which has the same design as Cavity2, by the Pound-Drever-Hall (PDH) method [32]. The frequency of the input laser beam is shifted by a double-pass acousto-optic modulator (AOM) that is driven by a voltage-controlled oscillator (VCO); thus it can be

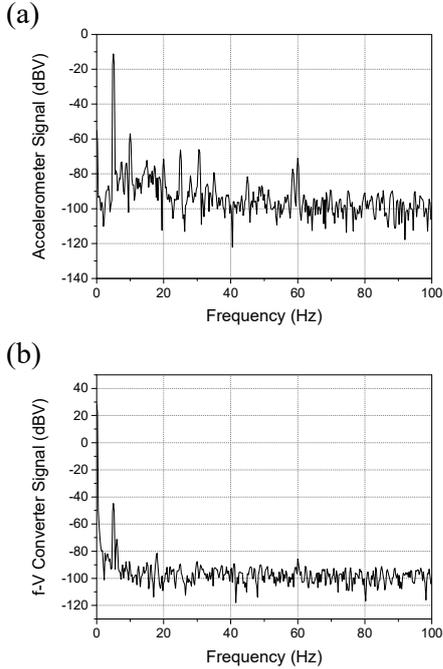

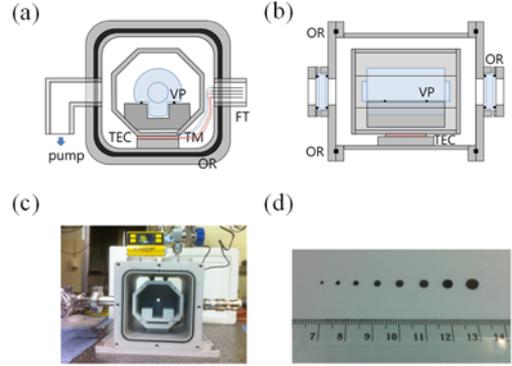

Figure 6. (a) Front view of the vacuum chamber for the vibration-sensitivity measurement: TEC, thermoelectric cooler; TM, thermistor; FT, 4-pin feed-through; VP, Viton pad; OR, O-ring seal. (b) Side view of the vacuum chamber. (c) Picture of the cavity installed in the partly assembled vacuum chamber. (d) Picture of the Viton support pads produced by hole punches.

Figure 5. Synchronous measurement of (a) the signal from the accelerometer and (b) the signal from the f-V converter, using a spectrum analyzer.

considered an independent laser source, compared to the original input. Then we used a PDH setup to stabilize the laser frequency to a resonance of either Cavity1 or Cavity2, as in Fig. 4. The optical cavity was contained in a vacuum chamber. The PDH error signal was fed back to the VCO for the stabilization. The frequency change due to vibration was measured by the beat note between this stabilized laser and the original input laser. The frequency of this beat-note signal was converted into a voltage using a frequency-to-voltage converter. The vacuum chamber and optical setup were placed on an active vibration-isolation table (AVI) (TS-300/LP, The Table Stable Ltd.). The AVI was used in "shaker" mode to provide a sinusoidal acceleration in one of the three orthogonal directions, to measure the vibration sensitivity. The magnitude of the acceleration was measured by a three-axis accelerometer. The frequency of the sinusoidal acceleration was chosen to be 5 Hz, under conditions such that the cavity deformation could be considered to be quasistatic, and the cross coupling among the vertical acceleration and two horizontal (transverse and longitudinal) accelerations was suppressed by more than 20 dB. The magnitudes of the acceleration and the frequency-to-voltage converter's signal were measured simultaneously, using a spectrum analyzer and by measuring the signals at 5 Hz (Figure

5). The cavity was supported with four Viton pads (Figs. 6(a) and (b)). Pads with various diameters between 1 mm and 5 mm were made using hole punches and 0.5-mm-thick Viton sheets. A picture of the produced Viton pads is shown in Fig. 6(d).

Since we have to open the vacuum chamber frequently to measure the vibration sensitivity at various support points with many kinds of Viton pads, the vacuum chamber's covers were sealed with O-rings (Figs. 6(a) and (b)). The vacuum chamber was made of aluminum, for convenient temperature control by a thermoelectric cooler and a thermistor, which would have advantages in finding the temperature of zero thermal expansion coefficient for the ULE cavity. The size of the vacuum chamber was 188 mm × 188 mm × 204 mm. The windows have coatings that are antireflective at 578 nm. A picture of the vacuum chamber with a cavity is given in Fig. 6(c).

## IV. RESULTS

The experimental results for the vibration sensitivity to vertical acceleration for Cavity1 and Cavity2 are shown in Figs. 2(a) and 2(b) respectively, together with the results of numerical simulation with the fixed model. The frequency change was normalized to the optical frequency. Four kinds of Viton pads, with various diameters and thicknesses, were used in the experiments. The vibration sensitivities were measured at four $z$ positions for each cavity. Each vibration sensitivity was obtained by averaging 10 measurements, and the error bars represent the standard deviations. As can be seen in Fig. 2, the experimental results agree well with the numerical simulations with a small, fixed support area. These experimental results are also shown with the results of the numerical simulations with the sliding model in Fig. 3, and they agree well with each other. These results can be interpreted such that the sliding model seems to be more appropriate, considering that the Viton rubber can be deformed easily, because it is mechanically soft. This is also consistent with the simulation results for the fixed model in Fig. 2, because the smaller the support area is, the more similar the conditions are to those of the sliding model. If we assume an experimental uncertainty of 0.5 mm in locating the $z$ position for zero vertical vibration sensitivity, it can be concluded that the vertical vibration sensitivities are less than $7\times10^{-12}/(m/s^2)$ and $9\times10^{-12}/(m/s^2)$ respectively for Cavity1 and Cavity2.

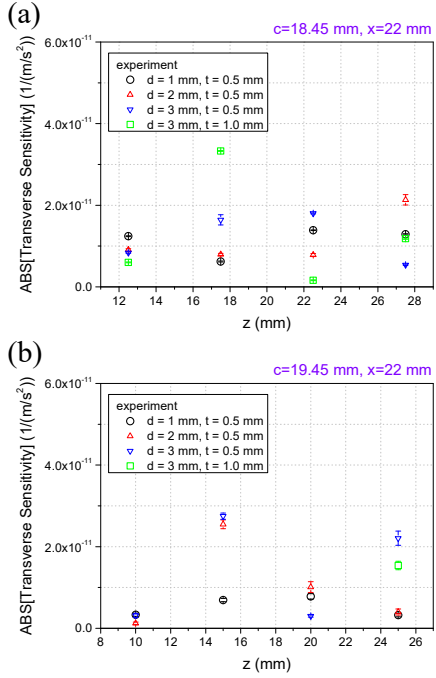
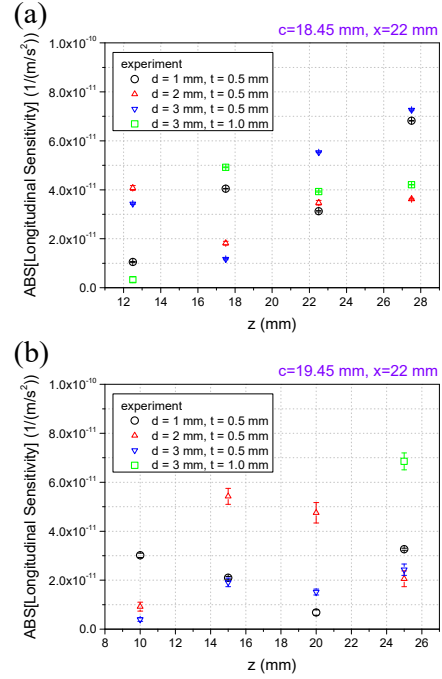

Figure 7. (a) Experimentally measured values of the transverse vibration sensitivity, for various support diameters and thicknesses. $c$ and $x$ were fixed at 18.45 and 22 mm respectively. (b) The same dependence as in (a), when only the value of $c$ was changed to 19.45 mm..

Figure 8. (a) Experimentally measured values of the longitudinal vibration sensitivity, for various support diameters and thicknesses. $c$ and $x$ were fixed at 18.45 and 22 mm respectively. (b) The same dependence as in (a), when only the value of $c$ was changed to 19.45 mm..

We also measured the horizontal (transverse and longitudinal) vibration sensitivities for both cavities. The transverse direction is along the $x$ displacement and the longitudinal is along the $z$ displacement in Fig. 1(c). We could measure only the magnitude of the horizontal vibration sensitivities, because the small signal for the frequency change made it difficult to determine the sign. The vibration-sensitivity measurements for acceleration in the transverse direction are shown in Fig. 7; most of the values were less than $2\times10^{-11}/(m/s^2)$ for both cavities. The vibration-sensitivity measurements for the acceleration in the longitudinal direction are shown in Fig. 8. The distribution of the longitudinal vibration-sensitivity values was more scattered; however, they were mostly less than $5\times10^{-11}/(m/s^2)$.

To explain the scattering of the vibration sensitivities, we performed a repeatability test of the measurement. Each of the data points in Fig. 9 was obtained by averaging 10 measurements, and the error bars represent the standard deviations. We obtained five vibration-sensitivity values for each direction under the same conditions (Cavity2, $d = 3.0$ mm, $t = 0.5$ mm, $z = 20$ mm, $x = 22$ mm), but with a new installation of the cavity each time. The vertical sensitivity had an average value of $2.9\times10^{-11}/(m/s^2)$ with a standard deviation of $5.9\times10^{-12}/(m/s^2)$. The transverse sensitivity had an average value of $1.2\times10^{-11}/(m/s^2)$ with a standard deviation of $1.0\times10^{-11}/(m/s^2)$. The longitudinal sensitivity had an average value of $3.1\times10^{-11}/(m/s^2)$ with a standard deviation of $1.8\times10^{-11}/(m/s^2)$. These results are consistent with the scattering of the vibration sensitivities in Figs. 2, 3, 7, and 8; thus it can be concluded that the scattering of the vibration-sensitivity measurements was due to the experimental repeatability of the cavity's support location.

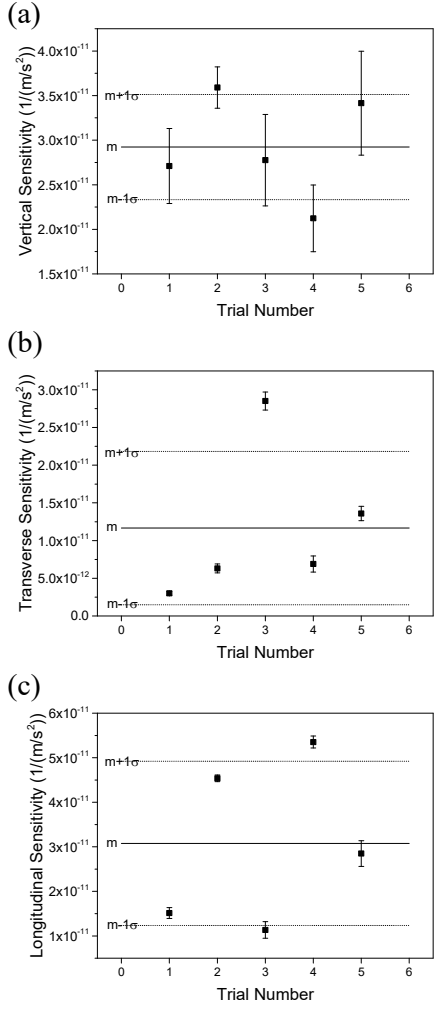

Figure 9. Repeatability of the measurement of vibration sensitivity. The solid and dotted lines indicate the values of the mean $m$ and standard deviation $\sigma$ for five measurements. (a) Vertical sensitivity: $m = 2.9\times10^{-11}$, $\sigma = 5.9\times10^{-12}$. (b) Transverse sensitivity: $m = 1.2\times10^{-11}$, $\sigma = 1.0\times10^{-11}$. (c) Longitudinal sensitivity: $m = 3.1\times10^{-11}$, $\sigma = 1.8\times10^{-11}$.

## V. CONCLUSION

We investigated the support-area dependence of vibration sensitivities of optical cavities, by numerical simulation and by experiment. Following the simulation results, a new cavity was designed which has a vibration-insensitive support position independent of the support's area. The numerical simulation was performed with two models: one with total constraint on the support area, and the other with only vertical constraint. The experimental results agreed well with the model based on vertical constraint. The vibration sensitivities in the three orthogonal directions (vertical, transverse, and longitudinal directions) were measured. The scattering of the sensitivity measurements could be explained by the repeatability in the location of the cavity's support.

frequency stabilization using an optical resonator," Appl. Phys. B **31**, 97-105 (1983).